\documentclass[aps,prd,superscriptaddress,showpacs,amsmath,amssymb]{revtex4}
\usepackage{amsmath}
\usepackage{amsfonts}
\usepackage{mathrsfs}
\usepackage{pstricks}
\usepackage{color}
\usepackage{graphicx}
\usepackage{slashed}
\usepackage{amssymb}

\DeclareMathOperator{\sgn}{sgn}
\newcommand{\be}{\begin{equation}}
\newcommand{\ee}{\end{equation}}
\newcommand{\ba}{\begin{array}{c}}
\newcommand{\ea}{\end{array}}
\newcommand{\bqa}{\begin{eqnarray}}
\newcommand{\eqa}{\end{eqnarray}}

\makeatletter
    
    \newcommand{\Rmnum}[1]{\expandafter\@slowromancap\romannumeral #1@}
    \makeatother

\begin{document}

\bibliographystyle{unsrt}

\title{\bf Meson-meson scattering in two-dimensional QCD}

\author{Guo-Ying Chen~\footnote{chengy@pku.edu.cn}}
\affiliation{Department of Physics and Astronomy, Hubei
University of Education, Wuhan 430205, China \vspace{0.2cm}}

\author{Yingsheng Huang~\footnote{huangys@ihep.ac.cn}}
\affiliation{Institute of High Energy Physics, Chinese Academy of Science, Beijing 100049,
China\vspace{0.2cm}}
\affiliation{School of Physics, University of Chinese Academy of
 Sciences, Beijing 100049, China\vspace{0.2cm}}

\author{Yu Jia~\footnote{jiay@ihep.ac.cn}}
\affiliation{Institute of High Energy Physics, Chinese Academy of Science, Beijing 100049,
China\vspace{0.2cm}}
\affiliation{School of Physics, University of Chinese Academy of
 Sciences, Beijing 100049, China\vspace{0.2cm}}

\author{Rui Yu~\footnote{yurui@ihep.ac.cn}}
\affiliation{Institute of High Energy Physics, Chinese Academy of Science, Beijing 100049,
China\vspace{0.2cm}}
\affiliation{School of Physics, University of Chinese Academy of
 Sciences, Beijing 100049, China\vspace{0.2cm}}

\date{\today}

\begin{abstract}
We extend the formalism pioneered by Callan, Coote and Gross to investigate the meson-meson scattering within the framework of
't Hooft model, {\it i.e.}, the two-dimensional QCD in the $N_c\to \infty$ limit.
We derive the analytic expressions for various two-body meson-meson
scattering amplitudes, concentrating on those quark diagrams which may be identified as the
meson-meson contact interaction vertex in the context of the
mesonic effective lagrangian in $1/N_c$ expansion.
We also carry out a detailed numerical study for the
meson-meson scattering for various quark flavors, and observe the near-threshold enhancement
in some channels. This may be viewed as the hint of the existence of the
tetra-quark state below two-meson threshold.
\end{abstract}


\maketitle

\section{Introduction}

The idea about the existence of the exotic hadrons, such as
tetraquark or pentaquark states, is as old as the naive quark model.
In recent years, this idea has actively revived as a dozen of new resonances were established experimentally,
some of which seem not to fit in the conventional $q\bar q$ or $qqq$ states (for a recent review, see
Ref.~\cite{Chen:2016qju,Guo:2017jvc}).
Most newly observed resonances are closely tied with the charmonium family,
generally referred to as the $XYZ$ states.
Some of them are considered as the viable candidates for the tetraquark or hadronic molecule.

The tetraquark states are usually studied within phenomenological models such as
the QCD sum rules or diquark model~\cite{Esposito:2014rxa,Esposito:2016noz}.
Unfortunately, the connection between these phenomenological approaches and the first principles of
QCD appears to be obscure. Recently, the \texttt{LHCb} experiment has discovered the long-awaited
doubly-charm baryon $\Xi_{cc}^{++}=ccu$~\cite{Aaij:2017ueg}.
Inspired by this important discovery, and with the guidance of heavy quark symmetry,
there have been convincing theoretical arguments that the stable
doubly-beauty tetraquark states,  as exemplified by the $bb \bar{u}\bar{d}$,
must exist~\cite{Karliner:2017qjm,Eichten:2017ffp}.

The $1/N_c$ expansion has historically served an influential nonperturbative tool of QCD~\cite{'tHooft:1973jz}.
This approach can successfully capture some gross traits of hadron phenomenology,
for instance the OZI rule and Regge behavior~\cite{Witten:1979kh}.
In the $N_c\to \infty $ limit, the QCD dynamics is dictated by the planar diagrams,
and one can show that all the mesons are stable and non-interacting with each other in the limit of
infinite number of color.
In fact, the meson-meson scattering first starts at order $1/N_c$.
In his famous series of Erice lectures, Coleman claimed
that the quark correlators possessing the tetraquark quantum number
make meson pairs and nothing else, as the connected tetraquark diagrams
are relatively $1/N_c$ suppressed~\cite{Coleman}.
Consequently there arises no nontrivial tetraquark
state in the large-$N_c$ limit. However, in 2013 Weinberg~\cite{Weinberg:2013cfa}
scrutinized Coleman's argument and pointed out some loophole.
Weinberg argued that the relatively $1/N_c$ suppression does not necessarily rule out the
existence of the tetraquark. He concluded that the existence of a {\it narrow} tetraquark is not
incompatible with large-$N_c$ QCD (for some further development along this direction, see
for instance \cite{Knecht:2013yqa,Cohen:2014tga,Maiani:2016hxw,Lucha:2017gqq}).

It is natural to speculate how to validate Weinberg's tetraquark state from the phenomenological angle.
It appears most appealing to search for these states by examining the meson-meson scattering within certain energy
range. These states may show up as a Breit-Wigner peak or manifest themselves through
some near-threshold enhancement on line-shape.

Albeit being qualitatively successful, the $1/N_c$ expansion can hardly
make any concrete quantitative prediction in the $3+1$-dimensional QCD.
Nevertheless, since the renowned work by 't Hooft in 1974~\cite{tHooft:1974pnl},
it becomes widely known that QCD
in the $1+1$ spacetime dimension (hereafter the 't Hooft model) is a solvable model of great value,
which mimics the realistic QCD in many aspects,
such as the color confinement, Regge behavior, chiral symmetry breaking and so on.
The 't Hooft model can be viewed as a fruitful theoretical laboratory to
test many interesting ideas in realistic
QCD.
It is the very goal of this paper to carry out a systematic study of the meson-meson scattering in
the 't Hooft model, with the particular incentive of searching for Weinberg's tetraquark state.
In an influential work by Callan, Coote and Gross~\cite{Callan:1975ps},
the theoretical framework of computing the
meson decay amplitude has been laid down using the formalism of the Bethe-Salpeter equation.
We will closely follow the recipe outlined in \cite{Callan:1975ps}, and extend
their work to the situation for the meson-meson scattering.
It is our hope that our result may shed some light on hunting the possible
tetraquark states in realistic QCD.

We remark that the meson-meson scattering has already been analyzed
within the 't Hooft model by Batiz, Pena and Stadler more than a decade ago ~\cite{Batiz:2003jm}.
Those authors claim to discover a Breit-Wigner peak, which is interpreted as
the a $\sigma$-like tetraquark state. Unfortunately, the authors of
\cite{Batiz:2003jm} appear to neglect some important class of Feynman diagrams also of the order $1/N_c$,
and consequently, their expressions are in fact gauge-dependent and sensitive to the
infrared regulator. Therefore, we feel obligated to revisit the meson-meson scattering
in the 't Hooft model from more consistent approach,
and consider all possible types of flavor textured possessed by
the incident and outgoing mesons.

In the next-to-leading order in $1/N_c$ expansion, the relevant Feynman diagrams for
meson-meson scattering include all the planar diagrams with the
quark line in the edges. As advocated by Witten~\cite{Witten:1979kh},
the equivalent description of $1/N_c$ expansion is to treat the meson as the effective degrees of freedom.
In this language, the two-body meson scattering process can be
classified into two classes of diagrams at tree level.
One type is composed of the meson exchange diagram, the other involves a single contact interaction vertex.
While the intermediate state of the $s$-channel meson
exchange diagram only contains an ordinary $q\bar q$ resonance,
the latter type of diagram may well accommodate a compact tetraquark structure.
Therefore, we will simply suppress those meson exchange diagrams, and
concentrate on the contact interaction diagrams to search for the exotic states.
The numerical studies reveal that we do not observe any Breit-Wigner resonance,
in contradiction with what is found in \cite{Batiz:2003jm}.
Nevertheless, we do observe
the near-threshold enhancement in the contact interaction amplitude
in some scattering channels. We tend to suggest that this
near-threshold enhancement may indicate the existence of
some tetraquark structure below the threshold.

The rest of the paper is structured as follows.
In Sec.~\Rmnum{2}, we recapitulate the essential ingredient of
the 't Hooft model, and review the formalism developed in \cite{Callan:1975ps}
on quark-antiquark scattering amplitude.
In Sec.~\Rmnum{3} we rederive the decay amplitude for a meson to two mesons,
within the framework of Callan, Coote and Gross.
In Sec.~\Rmnum{4}, following the recipe of \cite{Callan:1975ps},
we derive the analytic expressions for the contact
interaction amplitude affiliated with the meson-meson scattering with different flavors.
In Sec.~\Rmnum{5} we present our numerical results.
We summarize in Sec.~\Rmnum{6}.
In appendix~A, we describe some useful light-cone kinematics.
IN appendix~B, we enumerate the expressions for the contact
interaction amplitudes with all possible flavor structures.

\section{Quark-anti quark scattering in 't Hooft model}

\begin{figure}[hbt]
\begin{center}
  \includegraphics[width=10cm]{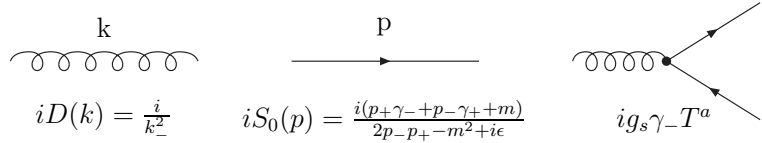}\\
  \caption{Feynman rules of ${\rm QCD}_2$ in the light-cone gauge.}\label{Feynmanrules}
  \end{center}
\end{figure}

The 't Hooft model is the two-dimensional QCD where the number of colors
is taken to be infinity~\cite{tHooft:1974pnl}.
The ${\rm QCD}_2$ Lagrangian reads
\begin{eqnarray}
\mathcal{L}=-\frac{1}{4}G^{a}_{\ \mu\nu} G^{a\mu\nu}\ + \sum_f \bar
q_{f}(i\gamma^\mu D_\mu-m_f)q_f,
\end{eqnarray}
where the sum is extended over quark flavors, and
\begin{eqnarray}
G^{a}_{\mu\nu}&=& \partial_{\mu} A^{a}_{\nu}-\partial_\nu A^{a}_{\mu}+ig_s f^{abc}A^b_{\mu} A^c_{\nu},\nonumber\\
D_\mu&=&\partial_\mu -ig_s A^a_{\mu} T^a,\nonumber\\
a&=&1,2,\ldots,N_c^2-1, \qquad f= u,d,s,c,b
\end{eqnarray}
The Lorentz indices $\mu,\nu$ run from $0$ to $1$.
$T^a$ are the $SU(N_c)$ generators, normalized as
$tr\left[T^aT^b\right]=\frac{1}{2}\delta^{ab}$, and $f^{abc}$ denotes the
structure constant. The quantization of ${\rm QCD}_2$
becomes particularly tractable if the light-cone gauge is imposed:
\begin{equation}
A_{-}=A^{+}=0,
\end{equation}
where
$A_{-}=\frac{1}{\sqrt{2}}(A^0+A^1)=\frac{1}{\sqrt{2}}(A_0-A_1)$.
A particular merit of the light-cone gauge is that
the non-ableian component of the field strength simply vanishes,
and the nonvanishing field
strength tensors are just
\begin{equation}
G_{+-}=-G_{-+}=-\partial_{-}A_{+},
\end{equation}
and the Lagrangian can then be written as
\begin{equation}
\mathcal{L}_{{\rm QCD}_2}= \frac{1}{2}\mbox{Tr}(\partial_{-}A_{+})^2+ \sum_f \bar{q}_f(i\partial_{+}\gamma_{-}+i\partial_{-}\gamma_{+}+{g_s}\gamma_{-}A_{+}-m_f)q_f.
\end{equation}
The light-cone representation for the Dirac $\gamma$ matrices obeys
\begin{equation}
\gamma^{+}=\frac{1}{\sqrt{2}}(\gamma^0\pm \gamma^1),\ \
(\gamma^+)^2=(\gamma^-)^2=0,\ \ \{\gamma^+,\gamma^-\}=2.
\end{equation}
In the light-cone gauge, there is neither occurrence of the ghost, nor the
physical (transverse) gluonic degrees of freedom. We
present the Feynman rules in the light-cone gauge in
Fig.~\ref{Feynmanrules}.

\begin{figure}[hbt]
\begin{center}
  \includegraphics[width=10cm]{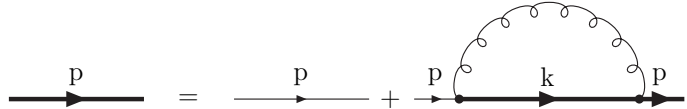}\\
  \caption{The Dyson-Schwinger equation for the quark self-energy. The thin line denotes the bare quark propagator
  and the solid line denotes the dressed quark propagator.}\label{dressedquark}
  \end{center}
\end{figure}

The quark self-energy diagrams satisfy the Dyson-Schwinger equation,
are depicted in Fig.~\ref{dressedquark}.
Notice diagrams with crossed gluons will be
suppressed by $1/N_c$, therefore the rainbow approximation becomes exact
in the large $N_c$ limit.
The Dyson-Schwinger equation then reads~\cite{tHooft:1974pnl,Batiz:2003jm}
\begin{equation}
S(p)=S_0(p)+i  \frac{N_c g^2_s}{2}
S(p)\left[\int\frac{d^2k}{(2\pi)^2}D(p-k)\gamma_{-}S(k)\gamma_{-}\right]S_{0}(p),
\end{equation}
where $S(p)$ denotes the dressed quark propagator. We assume $g_s\sim
\frac{1}{\sqrt{N_c}}$, so that $\frac{N_c g^2_s}{2}$ is kept fixed.
The solution to the above equation reads
\begin{eqnarray}
S(p)&=&\frac{p_{-}\gamma_{+}}{2p_{+}p_{-}-M^2-\frac{N_c
g_s^2}{2\pi}\frac{|p_{-}|}{\rho}+i\epsilon},\nonumber\\
M^2&=&m^2-\frac{N_c g_s^2}{2\pi},
\label{eq:bigM}
\end{eqnarray}
where $M$ denotes the mass of the dressed quark. $\rho$ is a dimensionful cutoff introduced to regularize
the infra-red divergence in the loop integral. For the
loop integral appearing in Fig.~\ref{dressedquark}, the value of $\rho$ is taken such that
$\rho<|k_{-}|<\infty$.

\begin{figure}[hbt]
\begin{center}
  \includegraphics[width=8cm]{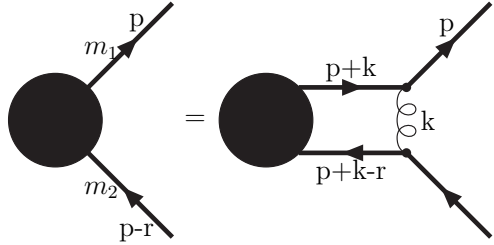}\\
  \caption{The Bethe-Salpeter equation for the $q\bar{q}$bound state.}\label{BSequation}
  \end{center}
\end{figure}

With the dressed quark propagator available, one then proceed to write the bound state equation
for the $q\bar q$ pair. In the large $N_c$ limit, the ladder approximation in Bethe-Salpeter equation
becomes exact, as shown in Fig.~\ref{BSequation}.
The corresponding Bethe-Salpeter equation reads
\begin{eqnarray}
\psi(p,r)&=&2iN_c g_s^2 p_{-}(p_{-}-r_{-})
\left[2p_{+}p_{-}-M_1^2-\frac{N_cg_s^2}{2\pi}\frac{|p_{-}|}{\rho}+i\epsilon\right]^{-1}\nonumber\\
&&\times \left[2(p_{+}-r_{+})(p_{-}-r_{-})-M_2^2-\frac{N_c
g_s^2}{2\pi}\frac{|p_{-}-r_{-}|}{\rho}+i\epsilon\right]^{-1}\nonumber\\
&&\times
\int\frac{d^2k}{(2\pi)^2}\frac{1}{k_{-}^2}\psi(p+k,r).\label{BS}
\end{eqnarray}
Defining $\varphi(p_{-},r)\equiv \int dp_{+}\psi(p,r)$, one then obtains
\begin{eqnarray}
\varphi(p_{-},r)&=&i\frac{N_c g_s^2}{2(2\pi)^2}\int dp_{+}
\left[p_{+}-\frac{M_1^2}{2p_{-}}-\frac{N_c
g_s^2}{4\pi}\frac{\sgn(p_{-})}{\rho}+i\epsilon\cdot \sgn(p_{-})\right]^{-1}\nonumber\\
&\times&\left[p_{+}-r_{+}-\frac{M_2^2}{2(p_{-}-r_{-})}-\frac{N_c
g_s^2}{4\pi}\frac{\sgn(p_{-}-r_{-})}{\rho}+i\epsilon\cdot \sgn(p_{-}-r_{-})\right]^{-1}\nonumber\\
&\times&\int dk_{-}\frac{\varphi(p_{-}+k_{-},r)}{k_{-}^2}.
\end{eqnarray}
Completing the $p_{+}$ integral and using
\begin{equation}
\int
dk_{-}\frac{\varphi(p_{-}+k_{-},r)}{k_{-}^2}=\frac{2}{\rho}\varphi(p_{-},r)
+P\int dk_{-}\frac{\varphi(p_{-}+k_{-},r)}{k_{-}^2},
\end{equation}
where
$P\frac{1}{k_{-}^2}=\frac{1}{2}(\frac{1}{(k_{-}+i\epsilon)^2}+\frac{1}{(k_{-}-i\epsilon)^2})$
indicates a principle-value prescription,
one then finds
\begin{eqnarray}
&&\left[r_{+}-\frac{M_2^2}{2(r_{-}-p_{-})}-\frac{M_1^2}{2p_{-}}-\frac{N_c
g_s^2}{2\pi \rho}+i\epsilon\right]\varphi(p_{-},r)\nonumber\\
&=&-\frac{N_c
g_s^2}{4\pi}\theta(p_{-})\theta(r_{-}-p_{-})\times\left[\frac{2}{\rho}\varphi(p_{-},r)+P\int
dk_{-}\frac{\varphi(p_{-}+k_{-},r)}{k_{-}^2}\right].
\end{eqnarray}
Clearly, the infra-red singularities in both sides cancel with each
other. After multiplying the factor $\frac{4\pi}{N_c g_s^2}r_{-}$ onto both
sides of the above equation, and introducing the following symbols:
 \begin{equation}
\mu^2=\frac{4\pi r_{+}r_{-}}{N_c g_s^2},\qquad \alpha_{1,2}=\frac{2\pi M_{1,2}^2}{N_c
g_s^2},\qquad x=\frac{p_{-}}{r_{-}},\label{Dless}
\end{equation}
one then recovers the celebrated 't Hooft equation:
\begin{equation}
\mu^2
\varphi(x)=\left(\frac{\alpha_{1}}{x}+\frac{\alpha_{2}}{1-x}\right)\varphi(x)-P\int_0^1
dy\frac{\varphi(y)}{(x-y)^2}.\label{teq}
\end{equation}

The solution of the 't Hooft equation leads to discrete mass enginevalues
$\mu_n^2$ ($n=0,1,2,\cdots$) for color-singlet mesons.
The corresponding wave functions
$\varphi_n$ satisfy the completeness and orthogonality relations:
\begin{equation}
\sum_{n}\varphi_{n}(x)\varphi_{n}^{\ast}(x^\prime)=\delta(x-x^\prime),\
\ \ \ \ \int_0^1\varphi_{n}^{\ast}(x)\varphi_{m}(x)dx=\delta_{nm}.
\end{equation}

\begin{figure}[hbt]
\begin{center}
  \includegraphics[width=14cm]{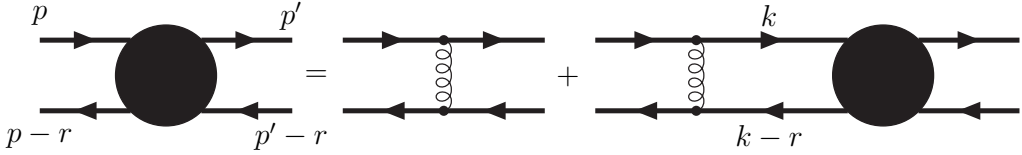}\\
  \caption{The Bethe-Salpeter equation for quark-antiquark scattering amplitude.}\label{quarkscattering}
  \end{center}
\end{figure}

In a similar vein, one may write down the Bethe-Salpeter
equation for the quark-antiquark scattering amplitude. As indicated
in Fig.~\ref{quarkscattering}, the corresponding
inhomogeneous Bethe-Salpeter equation reads~\cite{Callan:1975ps}
\begin{eqnarray}
\mathcal{T}(p,p^\prime;r)=-\frac{ig_s^2}{2(p_{-}-p_{-}^\prime)^2}
+i2N_c
g_s^2\int\frac{d^2k}{(2\pi)^2}\frac{1}{(k_{-}-p_{-})^2}\tilde{S}(k)\tilde{S}(k-r)\mathcal{T}(k,p^\prime;r),\label{quarkeq}
\end{eqnarray}
with $\tilde{S}(p)\gamma_+=S(p)$. This solution reads
\begin{eqnarray}
\mathcal{T}(x,x^\prime;r)
&=&-\frac{ig_s^2}{2r_{-}^2(x-x^\prime)^2}+\sum_{n}\frac{i}{r^2-r_{n}^2}\left\{\varphi_{n}(x)\frac{g_s^2}{2|r_{-}|}
\sqrt{\frac{N_c}{\pi}}\left[\theta(x(1-x))\frac{2|r_{-}|}{\rho}+\frac{\alpha_1}{x}+\frac{\alpha_2}{1-x}-\mu_{n}^2\right]\right\}\nonumber\\
&&\times\left\{\varphi_n^{\ast}(x^\prime)\frac{g_s^2}{2|r_{-}|}\sqrt{\frac{N_c}{\pi}}
\left[\theta(x^\prime(1-x^\prime))\frac{2|r_{-}|}{\rho}+\frac{\alpha_1}{x^\prime}+\frac{\alpha_2}{1-x^\prime}-\mu_n^2\right]\right\},\label{qqamp}
\end{eqnarray}
where $x=\frac{p_-}{r_-},\ \  x^\prime=\frac{p_-^\prime}{r_-}.$ The
amplitude bears infinite towers of poles located at $r^2=r_n^2,\ n=0,1,2\cdots$.
The physical interpretation of the above solution is clear, that the
summation of the $t$-channel multi-gluon exchange is equivalent to the
summation of the $s$-channel exchange of the quark-antiquark bound state. The
residue of the pole gives the meson-$q\bar{q}$ vertex function~\cite{Ball:1969}:
\begin{equation}
\Phi^{1,2}_{n}(x)=\varphi_{n}(x)\frac{g_s^2}{2|r_{-}|}
\sqrt{\frac{N_c}{\pi}}\left[\theta(x(1-x))\frac{2|r_{-}|}{\rho}+\frac{\alpha_1}{x}+\frac{\alpha_2}{1-x}-\mu_{n}^2\right].\label{Phi}
\end{equation}
The functions $\Phi^{1,2}_n(x)$ can be interpreted as the transition
amplitude between the meson and the quark-antiquark pair, which
serves an essential ingredient in our calculation for the meson-meson
scattering.

\section{Two-body strong decay of the meson}

\begin{figure}[hbt]
\begin{center}
  \includegraphics[width=8cm]{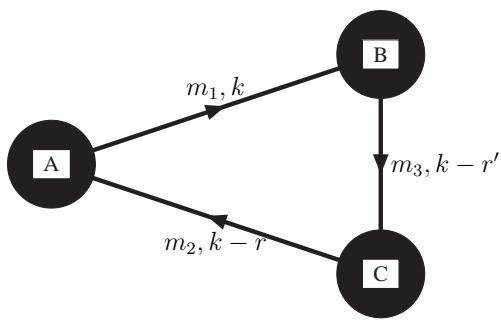}\\
  \caption{A two-body decay $A\rightarrow B+C$. $r$ is the incoming momentum of A, $r^\prime$ is the
  outgoing momentum of B, and $r^{\prime\prime}=r-r^\prime$ is the outgoing momentum of C.}\label{decay}
  \end{center}
\end{figure}

In this section, we take the two-body decay of the
meson in 't Hooft model as a warm-up exercise.
A mesonic two-body decay diagram is shown in
Fig.~\ref{decay}.
The decay amplitude can be written as
\begin{eqnarray}
i\mathcal{M}(A\to BC) &=&-iN_c\int\frac{d^2k}{(2\pi)^2}\frac{\Phi_{A}^{1,2}(x_A)\Phi_{B}^{1,3}(x_B)\Phi_{C}^{3,2}(x_C)}
{k_{+}-\frac{M_1^2}{2k_{-}}-\frac{N_c
g_s^2}{4\pi}\frac{\sgn(k_{-})}{\rho}+i\varepsilon\cdot
\sgn(k_{-})}\nonumber\\
&\times&\frac{1}{k_{+}-r_{+}-\frac{M_2^2}{2(k_{-}-r_{-})}-\frac{N_c
g_s^2}{4\pi}\frac{\sgn(k_{-}-r_{-})}{\rho}+i\varepsilon\cdot
\sgn(k_{-}-r_{-})}\nonumber\\
&\times&\frac{1}{k_{+}-r^\prime_{+}-\frac{M_3^2}{2(k_{-}-r^\prime_{-})}-\frac{N_c
g_s^2}{4\pi}\frac{\sgn(k_{-}-r^\prime_{-})}{\rho}+i\varepsilon\cdot
\sgn(k_{-}-r^\prime_{-})}.
\end{eqnarray}
where $r$ is the incoming momentum of particle $A$, and $r^\prime$
is the outgoing momentum of particle $B$. The arguments of the
$\Phi$ functions are defined as
\begin{equation}
x_A=\frac{k_{-}}{r_{-}},\ \ \ \ x_B=\frac{k_{-}}{r_{-}^\prime},\ \ \
\ x_{C}=\frac{k_{-}-r_{-}^\prime}{r_{-}-r_{-}^\prime}.
\end{equation}
One can first carry out the $k_+$ integral and take $\rho\rightarrow
0$ finally as the decay amplitude is infra-red safe. In doing that,
one should note that at least one of the $x_{A,B,C}$ can not lie in
the region $0<x<1$ due to the momentum conversation.  The final
expression for the decay amplitude reads
\begin{eqnarray}
i\mathcal{M}(A\to BC) &=&-g_s^2\sqrt{N_c/\pi}\left[\frac{1}{1-\omega}\int_{0}^{\omega}dx
\varphi_A\left(x\right)\varphi_B\left(\frac{x}{\omega}\right)\tilde{\Phi}_C(\frac{\omega-x}{\omega-1})\right.\nonumber\\
&&\left.-\frac{1}{\omega}\int_{\omega}^1dx
\varphi_A\left(x\right)\tilde{\Phi}_B(\frac{x}{\omega})\varphi_C\left(\frac{\omega-x}{\omega-1}\right)\right],
\end{eqnarray}
where we define $\omega=\frac{r_{-}^{\prime}}{r_{-}}$, $x=x_{A}$,
and $\tilde{\Phi}_{C,B}(x)=\int_0^1
dy\frac{\varphi_{C,B}(y)}{(x-y)^2}$.
This result has already been obtained by Barbon and companions~\cite{Barbon:1994au} long ago, which yet takes a different route,
{\it i.e.}, using the Hamiltonian and bosonization approach.
The numerical study of the various decay amplitudes have
also been conducted by Abdalla and collaborators~\cite{Abdalla:1998wv}.

\begin{figure}[hbt]
\begin{center}
  \includegraphics[width=10cm]{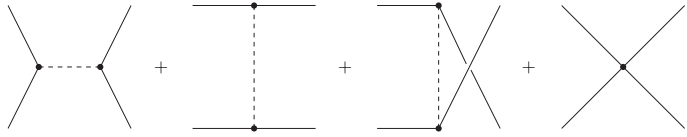}\\
  \caption{The tree diagrams for the scattering process $A+B\rightarrow C+D$,
  where the dashed line represents the exchanged meson. A sum over all species of mesons is understood. }\label{mesonscatteringaaa}
  \end{center}
\end{figure}

\section{The meson-meson scattering}

\begin{figure}[hbt]
\begin{center}
  \includegraphics[width=12cm]{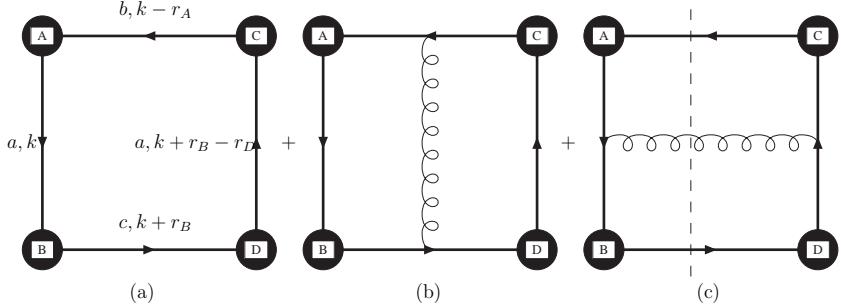}\\
  \caption{Four-body contact interaction part for $A(a\bar
b)+B(c\bar{a})\rightarrow C(a\bar{b})+D(c\bar{a})$. $r_A,r_B$ are
the incoming momenta of A and B respectively, and $r_C,r_D$ are the
outgoing momenta of C and D respectively. The dashed line is the cut
line.}\label{mesonscatteringa}
  \end{center}
\end{figure}

In this section, we will derive the analytical results for the
meson-meson scattering in the 't Hooft model. Consider the
meson-meson scattering $A+B\rightarrow C+D$, as Witten has
illustrated in Ref.~\cite{Witten:1979kh}, in the large $N_c$ limit
the leading contribution comes from the tree diagrams as shown in
Fig.~\ref{mesonscatteringaaa}. These tree diagrams can be classified
into two types, the contact-interaction type and the meson-exchange
type. To look for the exotic structure, we focus on the contact
interaction type diagrams in this work, because the meson exchange
diagrams contain only ordinary $q\bar q$ mesons. To figure out the
contact interaction amplitude in the t' Hooft model, one needs to
specify the flavor structure in the scattering. Let's first consider
the scattering which contain three different flavors
$A(a\bar{b})+B(c\bar{a})\rightarrow C(a\bar{b})+D(c\bar{a})$(where
$a,b,c$ denotes the quarks' flavors ). At the leading order, i.e.,
order $1/N_c$, there are infinite Feynman diagrams. We show three of
them in Fig.~\ref{mesonscatteringa}, which are the box diagram form
by the quark lines and the box diagrams with additional one gluon
exchange. The black bubble in Fig.~\ref{mesonscatteringa} represents
the meson-$q\bar q$ vertex function $\Phi^{q,\bar q}_n(x)$, thus
diagrams with gluon exchange between adjacent quark lines are also
included. Other diagrams which are also at the leading order and not
shown in Fig~\ref{mesonscatteringa} are those with multi-gluon
exchange in ladder fashion between nonadjacent quark lines. As
addressed in the above, the summation of the multi-gluon exchange
diagrams is equivalent to the summation of the $q\bar q$ meson
exchange diagrams, thus the sum of the infinite multi-gluon exchange
diagrams can be converted to the sum of the meson exchange diagrams
as shown in Fig.~\ref{mesonscatteringaaa}(one can refer to
Ref.~\cite{Callan:1975ps} for more detail). Therefore we only have
to consider the diagrams in Fig.~\ref{mesonscatteringa}, as their
sum equals to the contact interaction term. We take
Fig.~\ref{mesonscatteringa}(a) as an example to show some of the
details in our calculations. The amplitude for
Fig.~\ref{mesonscatteringa}(a) reads

\begin{eqnarray}
i\mathcal{M}_{box}&=&N_c\int\frac{d^2k}{(2\pi)^2}\frac{\Phi_{A}^{a,b}(x_A)\Phi_B^{c,a}(x_B)\Phi_C^{a,b}(x_C)\Phi_D^{c,a}(x_D)}
{k_{+}-\frac{M_a^2}{2k_{-}}-\frac{N_c
g_s^2}{4\pi}\frac{\sgn(k_{-})}{\rho}+i\varepsilon\cdot
\sgn(k_{-})}\nonumber\\
&\times&\frac{1}{k_{+}+r_{B+}-\frac{M_c^2}{2(k_{-}+r_{B-})}-\frac{N_c
g_s^2}{4\pi}\frac{\sgn(k_{-}+r_{B-})}{\rho}+i\varepsilon\cdot
\sgn(k_{-}+r_{B-})}\nonumber\\
&\times&\frac{1}{k_{+}+r_{B+}-r_{D+}-\frac{M_a^2}{2(k_{-}+r_{B-}-r_{D-})}-\frac{N_c
g_s^2}{4\pi}\frac{\sgn(k_{-}+r_{B-}-r_{D-})}{\rho}+i\varepsilon\cdot
\sgn(k_{-}+r_{B-}-r_{D-})}\nonumber\\
&\times&
\frac{1}{k_{+}-r_{A+}-\frac{M_b^2}{2(k_{-}-r_{A-})}-\frac{N_c
g_s^2}{4\pi}\frac{\sgn(k_{-}-r_{A-})}{\rho}+i\varepsilon\cdot
\sgn(k_{-}-r_{A-})},
\end{eqnarray}
where
\begin{equation}
x_{A}=\frac{k_{-}}{r_{A-}},  \ \ \
x_{B}=\frac{k_{-}+r_{B-}}{r_{B-}}, \ \ \
x_{C}=\frac{k_{-}+r_{B-}-r_{D-}}{r_{C-}}, \ \ \
x_{D}=\frac{k_{-}+r_{B-}}{r_{D-}}.
\end{equation}
We can first carry out the $k_+$ integral and expand the expression
in power of $\rho$, as we will postpone $\rho\rightarrow 0$ finally.
In doing the $k_+$ residual integral one should keep in mind that
$x_A$($x_C$) and $x_B$($x_D$) cannot lie in the region $0<x<1$
simultaneously, we can then find that Fig.\ref{mesonscatteringa}(a)
is of the order $\mathcal{O}(\rho)$. Therefore
Fig.\ref{mesonscatteringa}(a) gives vanishing contribution after
taking the limit $\rho\rightarrow 0$. One can easily check that
Fig.\ref{mesonscatteringa}(b) also gives vanishing contribution due
to the same reason. In contrast $x_A$($x_C$) and $x_B$($x_D$) can
lie in the region $0<x<1$ simultaneously in
Fig.~\ref{mesonscatteringa}(c), and this diagram gives nonvanishing
contribution. The difference between Fig.~\ref{mesonscatteringa}(c)
and the other two diagrams is that the S-channel cut line of this
diagram contains $q\bar q g$ state, while others contain only the
$q\bar q$ state. The final expression for
Fig.~\ref{mesonscatteringa} reads

\begin{eqnarray}
i\mathcal{M} &=&(1+\mathcal{C})i\mathcal{M}_0,\nonumber\\
i\mathcal{M}_0 &=&\theta(\omega_2-\omega_1)i2g_s^2\omega_1\int_0^1
dx\int_0^1
dy\frac{1}{(y\omega_1-\omega_2-x)^2}\varphi_A\left(\frac{\omega_2-\omega_1+x}{\omega_2-\omega_1+1}\right)\varphi_B\left(y\right)
\varphi_C\left(x\right)\varphi_D\left(\frac{y\omega_1}{\omega_2}\right),\nonumber\\
\end{eqnarray}
where $\omega_1=\frac{r_{B-}}{r_{C-}},
\omega_2=\frac{r_{D-}}{r_{C-}}$ and
\begin{equation}
\mathcal{C}=(A\leftrightarrow
C,\ \ B\leftrightarrow D,\ \ \omega_1\rightarrow
\frac{\omega_2}{1+\omega_2-\omega_1},\ \ \omega_2\rightarrow
\frac{\omega_1}{1+\omega_2-\omega_1}).
\end{equation}

\begin{figure}[hbt]
\begin{center}
  \includegraphics[width=10cm]{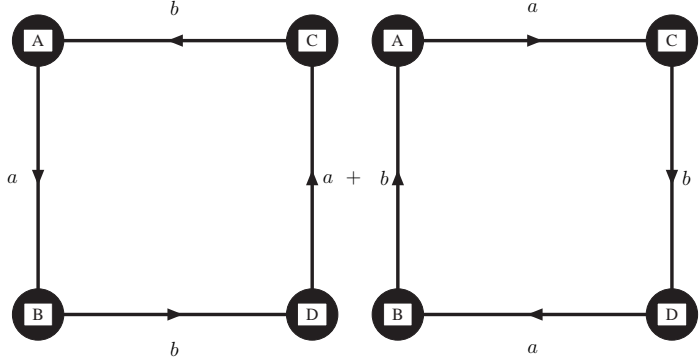}\\
  \caption{Box diagrams for $A(a\bar{b})+B(b\bar{a})\rightarrow C(a\bar{b})+D(b\bar{a})$.}\label{mesonssb}
  \end{center}
\end{figure}

The results for the four different flavor are similar. We then come
to study the meson-meson scattering for other flavor structures. We
find that there are more Feynman diagrams involved for meson-meson
scatterings with less flavor. The box diagrams for the two-flavor
scattering $A(a\bar{b})+B(b\bar{a})\rightarrow
C(a\bar{b})+D(b\bar{a}) $ are shown in Fig.~\ref{mesonssb}. One can
see that there are two box diagrams in the two-flavor scattering. To
calculate the contact interaction, the box diagrams with additional
one gluon exchange should also be included. Again, only diagrams
with S-channel cuts containing quark-gluon-anti quark states give
nonvanishing contribution. The final expression of the contact
interaction for $A(a\bar{b})+B(b\bar{a})\rightarrow
C(a\bar{b})+D(b\bar{a}) $ reads
\begin{equation}
i\mathcal{M}=(1+\mathcal{P})(1+\mathcal{C})i\mathcal{M}_{0},
\end{equation}
where $\mathcal{C}$ is defined above, and the operation
$\mathcal{P}$ is defined as $\mathcal{P}=(A\leftrightarrow B,\ \
C\leftrightarrow D,\ \ \omega_1\rightarrow
\frac{1+\omega_2-\omega_1}{\omega_2},\ \ \omega_2\rightarrow
\frac{1}{\omega_2})$.

\begin{figure}[hbt]
\begin{center}
  \includegraphics[width=12cm]{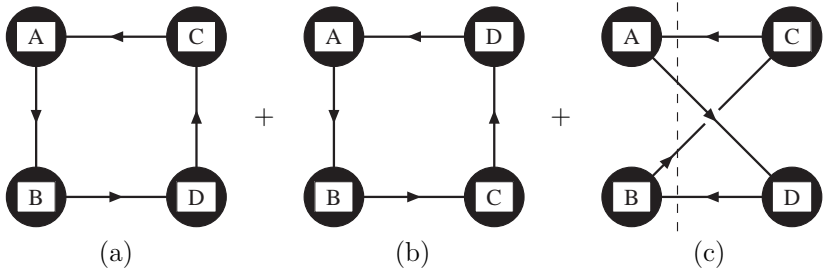}\\
  \caption{Three Box diagrams for $A(a\bar{a})+B(a\bar{a})\rightarrow C(a\bar{a})+D(a\bar{a})$.
  Other three diagrams are similar ones but with clockwise fermion loops.}\label{mesonssc}
  \end{center}
\end{figure}

For the single-flavor scattering $A(a\bar{a})+B(a\bar{a})\rightarrow
C(a\bar{a})+D(a\bar{a}) $, there are six box diagrams. We show three
of them in Fig.~\ref{mesonssc}, and others are corresponding
diagrams with clockwise fermion loops. To calculate the contact
interaction, we also need to consider the box diagrams with
additional one gluon exchange. Thus we need to consider 18 diagrams
in the single-flavor scattering. We note that while
Fig.~\ref{mesonssc}(a,b) give vanishing contribution,
Fig.~\ref{mesonssc}(c) gives nonvanishing contribution. This
difference is due to the fact that the two incoming particles $A$
and $B$ are directly connected by quark line in
Fig.~\ref{mesonssc}(a,b) but not in Fig.~\ref{mesonssc}(c). In other
words, all the S-channel cuts in Fig.~\ref{mesonssc}(c) contain
tetraquark states. We conclude that Feynman diagram with the
S-channel cut line containing only the $q\bar q$ state gives
vanishing contribution. Therefore, 10 of the 18 diagrams give
nonvanishing contributions. The calculation is tedious but
straightforward. The only subtlety is that the amplitude for
Fig.~\ref{mesonssc}(c) contains the divergent part
$\mathcal{O}(1/\rho)$, and the divergent part can be exactly
canceled by the contributions from the corresponding diagrams with
additional one gluon exchange~\cite{Callan:1975ps}. The final
expression for the contact interaction of
$A(a\bar{a})+B(a\bar{a})\rightarrow C(a\bar{a})+D(a\bar{a}) $ reads
\begin{eqnarray}
i\mathcal{M}
=(1+\mathcal{R})(1+\mathcal{P})(1+\mathcal{C})i\mathcal{M}_{0}+(1+\mathcal{R})i\mathcal{M}_{1},
\end{eqnarray}
\begin{figure}[b]
\begin{center}
  \includegraphics[width=7cm]{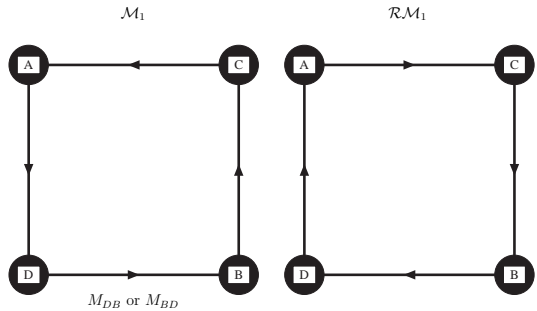}\\
  \caption{The Feynman diagrams for $\mathcal{M}_1$ and $\mathcal{RM}_1$, where $M_{DB}$($M_{BD}$) indicates the
  $M$ of the quark connecting $B$ meson and $D$ meson. }\label{M1_Feynman}
  \end{center}
\end{figure}
where
\begin{eqnarray}
i\mathcal{M}_{1}&=&-(1+\mathcal{Q})\theta(1-\omega_1)i2g_s^2\int_0^1
dx P\int_0^1
dy\frac{\omega_1\omega_2}{[(y-1)\omega_1+(1-x)\omega_2]^2}\varphi_A\left(\frac{x\omega_2}{1+\omega_2-\omega_1}\right)\varphi_B\left(y\right)
\varphi_C\left(y\omega_1\right)\varphi_D\left(x\right)\nonumber\\
&&-(1+\mathcal{C})\theta(\omega_2-\omega_1)i2g_s^2\int_0^1dx P\int_0^1
dy\frac{\omega_1}{(y\omega_1-x)^2}\varphi_A\left(\frac{x+\omega_2-\omega_1}{1+\omega_2-\omega_1}\right)\varphi_B\left(y\right)
\varphi_C\left(x\right)\varphi_D\left(\frac{(y-1)\omega_1+\omega_2}{\omega_2}\right)\nonumber\\
&&
-(1+\mathcal{Q}+\mathcal{P}+\mathcal{C})\theta(\omega_2-\omega_1)\theta(\omega_1-1)i\frac{4\pi}{N_c}\int_0^1
dx\left[2r_{C+}r_{C-}+2r_{D+}r_{C-}+\frac{M_{DB}^2}{x-\omega_1}+\frac{M_{CA}^2}{x-1}\right.\nonumber\\
&&\left.-\frac{M_{AD}^2}{x-\omega_1+\omega_2}
-\frac{M_{BC}^2}{x}\right]
\times\varphi_A\left(\frac{x-\omega_1+\omega_2}{1+\omega_2-\omega_1}\right)\varphi_B\left(x/\omega_1\right)
\varphi_C\left(x\right)\varphi_D\left(\frac{x-\omega_1+\omega_2}{\omega_2}\right),\nonumber\\
\end{eqnarray}
where $M_{DB}$ indicates the $M$, defined in (8), of the corresponding quark propagator connecting the meson $D$ and the meson $B$, so are the terms $M_{CA}$ , $M_{AD}$ and $M_{BC}$. We also have
\begin{equation}
\mathcal{R}=(C\leftrightarrow D,\ \ \omega_1\rightarrow
\frac{\omega_1}{\omega_2},\ \ \omega_2\rightarrow 1/\omega_2),\ \ \
\mathcal{Q}=(B\leftrightarrow C,\ \ A\leftrightarrow D,\ \
\omega_1\rightarrow 1/\omega_1,\ \ \omega_2\rightarrow
\frac{1+\omega_2-\omega_1}{\omega_1}).
\end{equation}
We have shown the Feynman diagrams of $\mathcal{M}_1$ and $\mathcal{RM}_1$ in Fig.\ref{M1_Feynman}. If there is an $\mathcal{R}$ acting on the $\mathcal{M}_1$, one should refer to the feynman diagram of $\mathcal{RM}_1$ to figure out the flavor of the $M$s.

For completeness, we also list the contact interaction terms for
meson-meson scatterings with other flavor structures in the
appendix. We would like to mention that parts of the analytical
results are also given in Ref.~\cite{Barbon:1994au}.

We end this section by commenting on the preceding calculation of meson-meson scattering by
Batiz {\it et al.}~\cite{Batiz:2003jm}.
One of the problems is that their Feynman rules seems not to distinguish
the outgoing quark and the incoming quark for a meson vertex. This
leads to a nonvanishing result for Fig.~\ref{mesonscatteringa}$(a)$,
which is vanishing in our paper. The other severe mistake is that they have
missed the one-gluon exchange diagrams. As mentioned before,
Fig.~\ref{mesonssc}$(c)$ possesses a term containing factor
$\frac{1}{\rho}$, which should be canceled by the corresponding
diagrams with an additional gluon exchange diagram. Thus
Fig.~\ref{mesonssc}$(c)$ alone is IR divergent. However,
since the authors of \cite{Batiz:2003jm} employed the principle-value as their default
IR regulator, they have not realized their results are actually IR divergent.
Therefore, their result for
Fig.~\ref{mesonssc}$(c)$ cannot be affiliated with physical significance.
We stress that, by confirming that our final result is free from the IR cutoff $\rho$,
provides a quite nontrivial consistency check for our calculation.

\section{Numerical results}

\begin{figure}[hbt]
\begin{center}
  \includegraphics[width=10cm]{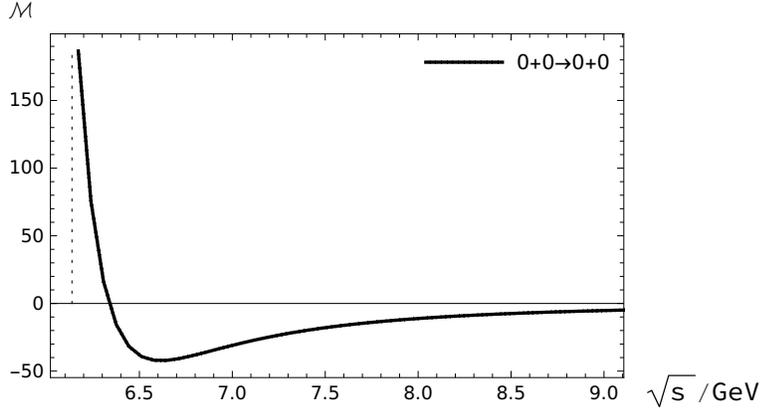}\\
  \caption{Amplitudes for the contact term in $A(c\bar c)+B(c\bar c)\rightarrow C(c\bar c)+D(c\bar c)$.}
  \label{the1flavor}
  \end{center}
\end{figure}

\begin{figure}[hbt]
\begin{center}
  \includegraphics[width=10cm]{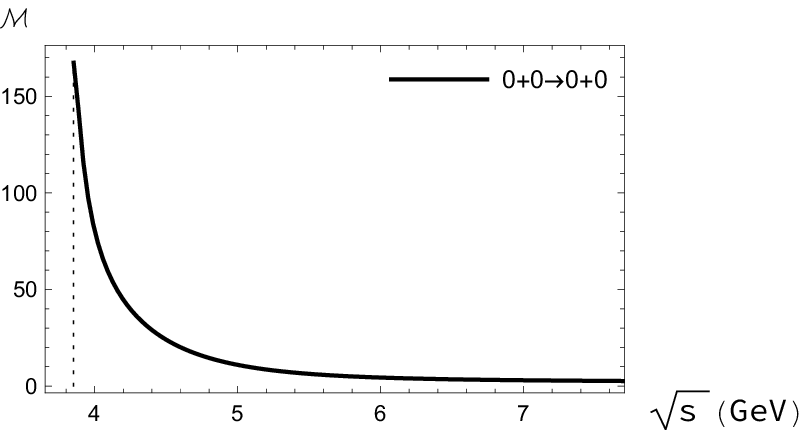}\\
  \caption{Amplitudes for the contact term in $A(c\bar s)+B(c\bar s)\rightarrow C(c\bar s)+D(c\bar s)$.}
  \label{nr2flavor}
  \end{center}
\end{figure}

\begin{figure}[hbt]
\begin{center}
  \includegraphics[width=10cm]{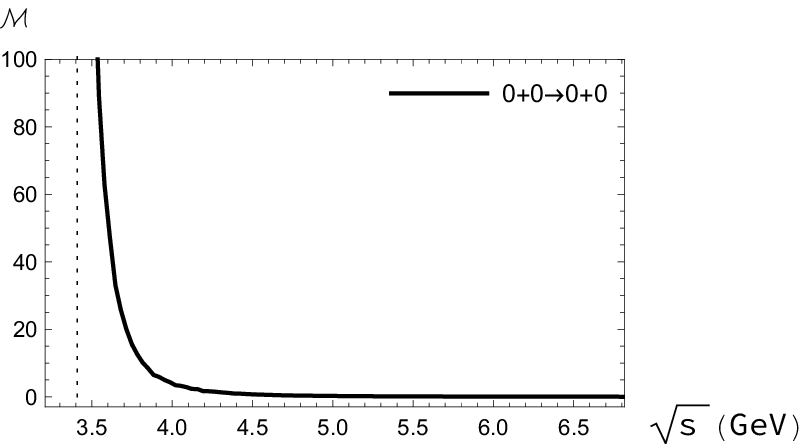}\\
  \caption{Amplitudes for the contact term in $A(c\bar u)+B(c\bar d)\rightarrow C(c\bar u)+D(c\bar d)$.}
  \label{nr1flavor}
  \end{center}
\end{figure}

\begin{figure}[hbt]
\begin{center}
  \includegraphics[width=10cm]{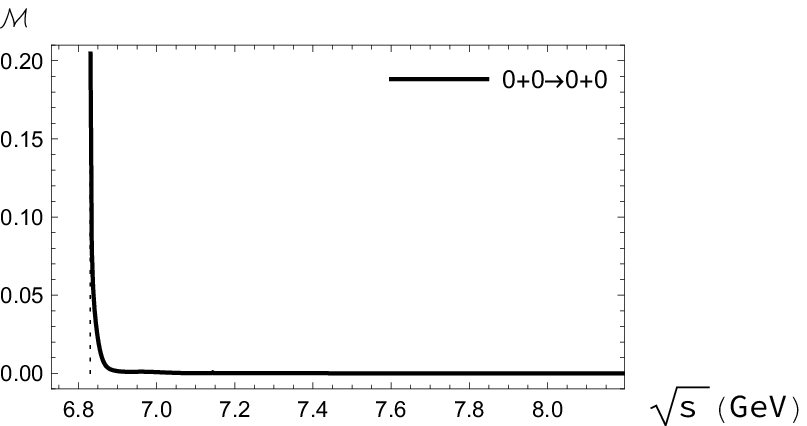}\\
  \caption{Amplitudes for the contact term in $A(c\bar d)+B(b\bar s)\rightarrow C(b\bar d)+D(c\bar s)$.}
  \label{nr3flavor}
  \end{center}
\end{figure}

We now move to the numerical study of the meson-meson scattering.
To evaluate the contact interaction amplitude of the meson-meson
scattering, we first need the numerical results for the meson light-cone wave
functions. These functions can be obtained by solving the 't Hooft
equation with the standard eigenvalue
routines~\cite{Lebed:2000gm,Brower:1978wm}. Following
Ref.~\cite{Jia:2017uul}, we express any dimensional quantity in
unit of $\sqrt{2\lambda}=340$ MeV, where $\lambda=\frac{g_s^2
N_c}{4\pi}$. To mimic the realistic meson spectrum in ${\rm QCD}_4$, the
bare quark masses are chosen as $m_{u}=15.3\textup{MeV},\ m_{d}=30.6\textup{ MeV},\ m_{s}=254\textup{MeV},\
m_{c}=1.44\textup{Gev}$~\cite{Jia:2017uul}, and $m_{b}=4.61\textup{Gev}$.
For the sake of completeness, we show the numerical
results for the one-flavor scattering $A(c\bar c)+B(c\bar c)\rightarrow C(c\bar c)+D(c\bar c)$ in Fig.~\ref{the1flavor},
the two-flavor scattering $A(c\bar s)+B(c\bar s)\rightarrow C(c\bar s)+D(c\bar s)$ in Fig.~\ref{nr2flavor},
the three-flavor scattering $A(c\bar u)+B(c\bar d)\rightarrow C(c\bar u)+D(c\bar d)$ in Fig.~\ref{nr1flavor},
and the four-flavor scattering $A(c\bar d)+B(b\bar s)\rightarrow C(b\bar d)+D(c\bar s)$ in Fig.~\ref{nr3flavor}.
For simplicity, we only consider the scattering of the ground-state mesons, which are simply represented by $0+0\rightarrow 0+0$.
From our numerical results, we do observe clear enhancement near the threshold.
Upon varying the bare quark mass, we find that the near threshold enhancement does not disappear.
We also find that this enhancement is not necessary a universal feature for meson-meson scattering.
For example, we do not observe the near-threshold enhancement in the channel
$A(c\bar d)+B(b\bar s)\rightarrow C(b\bar d)+D(c\bar s)$.

\section{Summary}

In summary, we have carried out a comprehensive study on the meson-meson scattering in the 't Hooft model.
Since the original goal is to search for the possible tetraquark state,
we intentionally only examine the contact interaction part of the
meson-meson scattering amplitude. We derive the analytic results for the corresponding amplitude,
considering all possible flavor structures. We find that only Feynman diagrams with
the $s$-channel cut on the $q\bar q g$ or $q\bar q q\bar q$ intermediate states can make
nonvanishing contribution. Reassuringly, we explicitly verify that
the contact interaction amplitude is free from the IR regulator $\rho$.
Our numerical study reveals that these diagrams may generate the near-threshold
enhancement for some channels of meson-meson scattering. This may be viewed as a sign
of  the existence of the tetraquark state below threshold.

\section*{ACKNOWLEDGMENTS}
 The work of Y.~J., Y.-S.~H. and R.~Y. is supported in part by the National Natural Science Foundation of China
under Grants No.~11875263, No.~11475188, No.~11621131001 (CRC110 by DFG and NSFC).

\section*{APPENDIX A: light-cone KINEMATICS}
In the $1+1$
dimensional case, there is only one kinematical degree of freedom
involved in a $2\to 2$ scattering within the center-of-mass frame.
Thus, in principle we can express all the results in terms of the
squared center-of-mass energy $s$. Nevertheless, we employ two
kinematical variables in our calculations for convenience, which are
defined as
\begin{align}
\label{def:omega12}
\omega_1=\frac{r_{B-}}{r_{C-}},\ \ \
\omega_2=\frac{r_{D-}}{r_{C-}}.\ \ \
\end{align}
with all the final results expressed by $\omega_1$ and $\omega_2$. To be clear, we also list the following equations
\begin{align}
\frac{r_{A-}}{r_{B-}}=\frac{1+\omega_2-\omega_1}{\omega_1},\ \frac{r_{D-}}{r_{C-}}=\omega_2,
\label{the_relations}
\end{align}
with the relations~\eqref{the_relations} and the light-cone dispersion relation $r_{X+}=\frac{M_X^2}{2r_{X-}}$, we can transform the following two equations
\begin{align*}
s=2(r_{C+}+r_{D+})(r_{C-}+r_{D-}),\\
s=2(r_{A+}+r_{B+})(r_{A-}+r_{B-}),
\end{align*}
into
\begin{subequations}
\label{eq:kinetic equations}
\begin{eqnarray}
s\frac{\omega_2}{1+\omega_2}&=&M_C^2\omega_2+M_D^2,\label{eq:kinetic equation a}\\
s\frac{\omega_2}{1+\omega_2}&=&M_A^2\frac{\omega_2}{1+\omega_2-\omega_1}+M_B^2\frac{\omega_2}{\omega_1}.\label{eq:kinetic equation b}
\end{eqnarray}
\end{subequations}
which show the relations between $s$ and $\omega$s.

To get real solutions for equations~\eqref{eq:kinetic equations}, one needs to put $s$ above the threshold $\max\left\{(M_A+M_B)^2,(M_C+M_D)^2\right\}$. Furthermore, once a suitable $s$ is selected, there will be four solutions for $(\omega_1,\omega_2)$. The relations between $\omega$s and the direction of mesons' momentums is shown in Table.\ref{the_direction}. Since for a
specific scattering the incoming momenta are fixed, we choose the first two lines of Table.~\ref{the_direction} in our calculation. Besides, it should be mentioned that when $C$ and $D$ are the same mesons, the first two lines of Table.\ref{the_direction} are equivalent since the meson $C$ and the meson $D$ are identical particles.

\begin{table}[tbh]
\begin{centering}
\begin{tabular}{cccccc}
\hline
$\omega_1$ & $\omega_2$ & A & B & C & D \tabularnewline
\hline
smaller & smaller & $\rightarrow$ & $\leftarrow$ & $\rightarrow$ & $\leftarrow$\tabularnewline
\hline
smaller & larger & $\rightarrow$ & $\leftarrow$ & $\leftarrow$ & $\rightarrow$\tabularnewline
\hline
larger & smaller & $\leftarrow$ & $\rightarrow$ & $\rightarrow$ & $\leftarrow$\tabularnewline
\hline
larger & larger & $\leftarrow$  & $\rightarrow$ & $\leftarrow$ & $\rightarrow$\tabularnewline
\hline
\end{tabular}
\par\end{centering}
\begin{centering}
\caption{The relation between $\omega$s and the directions of mesons' momentums $r^1$ in the center-of-mass frame, where $\rightarrow$($\leftarrow$) indicates a positive(negative) $r^1$. There are two solutions for each $\omega$, namely four groups of solutions. ``Smaller"(``larger") means that we choose the smaller(larger) $\omega_{1(2)}$.}
\label{the_direction}
\par\end{centering}
\end{table}

\section*{APPENDIX B: CONTACT INTERACTION TERMS FOR MESON-MESON SCATTERINGS WITH DIFFERENT FLAVOR STRUCTURES}

Contact interaction amplitudes for meson-meson scattering with different
flavor structures can be expressed with the functions
$\mathcal{M}_0$ and $\mathcal{M}_1$ defined in Sec.\Rmnum{3}.
\begin{itemize}
\item Contact interaction terms for meson-meson scatterings with four
different flavors:\\
\begin{eqnarray}
A(a\bar{d})+B(b\bar{a})\rightarrow C(c\bar{d})+D(b\bar
 c)&&:\ \ (1+\mathcal{C})\mathcal{M}_0,\nonumber\\
A(a\bar{d})+B(c\bar{b})\rightarrow C(c\bar{d})+D(a\bar
 b)&&:\ \ \mathcal{M}_1.
\end{eqnarray}

\item Contact interaction terms for meson-meson scatterings with
three different flavors:\\
$$\begin{array}{lllll}
&&A(a\bar{c})+B(b\bar{a})\rightarrow C(b\bar{c})+D(b\bar
 b)&&:\ \ (1+\mathcal{C})\mathcal{M}_0,\nonumber\\
&&A(a\bar{c})+B(b\bar{a})\rightarrow C(c\bar{c})+D(b\bar
 c)&&:\ \  (1+\mathcal{C})\mathcal{M}_0,\nonumber\\
&&A(a\bar{b})+B(c\bar{a})\rightarrow C(a\bar{b})+D(c\bar
 a)&&:\ \  (1+\mathcal{C})\mathcal{M}_0,\nonumber\\
&&A(a\bar{b})+B(b\bar{a})\rightarrow C(c\bar{b})+D(b\bar
 c)&&:\ \ (1+\mathcal{C})\mathcal{M}_0,\nonumber\\
&&A(a\bar{a})+B(b\bar{a})\rightarrow C(c\bar{a})+D(b\bar
 c)&&:\ \ (1+\mathcal{C})\mathcal{M}_0, \nonumber\\
&&A(a\bar{c})+B(a\bar{a})\rightarrow C(b\bar{c})+D(a\bar
 b)&&:\ \ (1+\mathcal{C})\mathcal{M}_0, \nonumber\\
&&A(a\bar{c})+B(b\bar{b})\rightarrow C(b\bar{c})+D(a\bar
 b)&&:\ \ \mathcal{M}_1, \nonumber\\
&&A(a\bar{c})+B(c\bar{b})\rightarrow C(c\bar{c})+D(a\bar
 b)&&:\ \ \mathcal{M}_1,\nonumber\\
&&A(a\bar{c})+B(a\bar{b})\rightarrow C(a\bar{c})+D(a\bar
 b)&&:\ \ \mathcal{M}_1,\nonumber\\
&&A(a\bar{c})+B(b\bar{c})\rightarrow C(b\bar{c})+D(a\bar
 c)&&:\ \ \mathcal{M}_1.\nonumber
\end{array}$$

\item Contact interaction terms for meson-meson scatterings with
two different flavors:\\
$$\begin{array}{lllll}
A(a\bar{a})+B(b\bar{a})\rightarrow C(b\bar{a})+D(b\bar
 b)&&:\ \ (1+\mathcal{C})\mathcal{M}_0,\nonumber\\
A(a\bar{a})+B(a\bar{a})\rightarrow C(b\bar{a})+D(a\bar
 b)&&:\ \
(1+\mathcal{R}\mathcal{P})(1+\mathcal{C})\mathcal{M}_0,\nonumber\\
A(a\bar{b})+B(b\bar{a})\rightarrow C(b\bar{b})+D(b\bar
 b)&&:\ \ (1+\mathcal{R})(1+\mathcal{C})\mathcal{M}_0,\nonumber\\
A(a\bar{b})+B(b\bar{a})\rightarrow C(a\bar{b})+D(b\bar
 a)&&:\ \ (1+\mathcal{P})(1+\mathcal{C})\mathcal{M}_0,\nonumber\\
A(a\bar{b})+B(a\bar{a})\rightarrow C(b\bar{b})+D(a\bar
 b)&&:\ \ (1+\mathcal{C})\mathcal{M}_0,\nonumber\\
A(a\bar{a})+B(b\bar{b})\rightarrow C(b\bar{a})+D(a\bar
 b)&&:\ \ \mathcal{M}_1,\nonumber\\
A(a\bar{b})+B(b\bar{a})\rightarrow C(b\bar{b})+D(a\bar
 a)&&:\ \ \mathcal{M}_1,\nonumber\\
A(a\bar{b})+B(a\bar{b})\rightarrow C(a\bar{b})+D(a\bar
 b)&&:\ \ (1+\mathcal{R})\mathcal{M}_1,\nonumber\\
A(a\bar{a})+B(b\bar{a})\rightarrow C(a\bar{a})+D(b\bar
 a)&&:\ \ (1+\mathcal{C})\mathcal{M}_0+\mathcal{R}\mathcal{M}_1,\nonumber\\
A(a\bar{b})+B(a\bar{a})\rightarrow C(a\bar{b})+D(a\bar{a})&&:\ \ (1+\mathcal{C})\mathcal{M}_0+\mathcal{M}_1.
\end{array}$$

\item Contact interaction terms for meson-meson scattering with
single flavor:\\
\begin{equation}
A(a\bar{a})+B(a\bar{a})\rightarrow C(a\bar{a})+D(a\bar
 a):\ \
(1+\mathcal{R})(1+\mathcal{P})(1+\mathcal{C})\mathcal{M}_0+(1+\mathcal{R})\mathcal{M}_1.\nonumber\\
\end{equation}

\end{itemize}

\par\par

\end{document}